\documentstyle[12pt]{article}

\newcommand{\be}{\begin{equation}}
\newcommand{\ee}{\end{equation}}
\newcommand{\ba}{\begin{eqnarray}}
\newcommand{\ea}{\end{eqnarray}}
\newcommand{\no}{\nonumber}

\everymath{\displaystyle}

\begin{document}
{\normalsize 

\begin{center}
{\bf Professor Nambu, String Theory and Moonshine Phenomenon}
\end{center}
\vskip1cm
\begin{center}
\bf {Tohru Eguchi}
\end{center}

\begin{center} 
{\sl Department of Physics and Research Center for Mathematica Physics}
\end{center}
\begin{center}
{\sl Rikkyo University, Tokyo 171-8501, Japan}
\end{center}

\begin{abstract}
 I first recall the last occasion of meeting the late Professor Yoichiro Nambu in a hospital in Osaka.
I then present a brief introduction to the moonshine phenomenon in string theory which is under recent investigations.  
\end{abstract}

\section{Farewell to Prof.Nambu}

It is  very sad that we  no longer see and talk to professor Nambu these days. As everybody knows, prof. 
Nambu has been one of the greatest theoretical physicists of our time. He was an idol of my generation. He contributed so much to transform and bring particle physics to the present day standard model. He further introduced the string theory which will be the main ingredient of particle physics of future generation. We no longer have the special privilege of having a great man around us.

Last time when I saw him was in the early June, 6/7  in a hospital of Osaka. This day   was  a Sunday and I wanted to come to Osaka and see prof.Nambu. I used to see him about once a month in these days. It  turned out that on the previous day, 6/6 he had a serious pain in his heart which looked like an after effect of a heart attack he suffered before. And he was kept in an ICU. 

I was not sure if I could see him but it turned out I was allowed to enter the ICU room. 
Prof.Nambu seemed recovered from the pain of the day before and was in a good condition and spirit. He was smiling with his family. When first he saw me, he mentioned a word "zero point life" pointing to himself. He was making a joke out of the health condition of himself. 

I think this was one of a few happy moments during his last fight against illness. I stayed for a few hours and went back to Tokyo. I want to keep the memory of this scene of ICU forever.     \\

Now let me turn to the discussion of physics. About five years ago together with my collaborators I have found some curious phenomenon in string theory, i.e. appearance of exotic discrete symmetry in the theory \cite{EOT}. This phenomenon is now called as Mathieu moonshine  and is under intensive study. Prof.Nambu was curious to hear about the 
 story, however, there was no chance to tell him the details.  Today I would like to give you a brief introduction  to moonshine  phenomena which possibly play  
 interesting role in string theory in the future.\\

Before going into the moonshine phenomenon in string theory let me briefly recall the story of monstrous moonshine which is very well-known.
Modular $J$ function has a $q$-series expansion 
\ba
&&J(q)={1\over q}+744+196884q+21493760q^2+864299970 q^3 \no \\
&&\hskip3cm +20245856256q^4+333202640600q^5 +\cdots\no
\ea
\ba
q=e^{2\pi i\tau}, \hskip2mm Im(\tau)>0,\hskip2mm J(\tau)=J({a\tau+b\over c\tau+d}), 
\hskip2mm \left(\begin{array}{cc} a & b\\ c & d \end{array}\right)\in SL(2,Z)\no
\ea                 
It turns out  q-expansion coefficients of $J$-function are decomposed into a sum of dimensions of irreducible  representations of the monster group  $M$ as
 \ba
&& \hskip-3mm 196884=1+196883,\hskip3mm 21493760=1+196883+21296876, \no \\
 && \hskip-3mm 864299970=2\times 1+2\times 196883+21296876+842609326, \no \\
  && \hskip-3mm 20245856256=1\times 1+3\times 196883+2\times 21296876 \no\\
  &&+842609326+19360062527, \cdots \no \ea     
 Dimensions of some irreducible representations of monster are in fact given by\\
 \hspace*{3cm} $\{1,\,196883,\,21296876,\,842609326,\\
 \,\hspace*{5cm}18538750076,\, 19360062527\cdots\}$  \\  
           
Monster group is the largest sporadic discrete group, of order  $\approx 10^{53}$ 
 and the   strange relationship between modular form and the largest discrete group was first noted by McKay.

To be precise we may write as 
\ba
&&J_1(\tau)=J(q)-744=\sum_{n=-1}c(n)q^n, \hskip2cm c(0)=0\no\\
&&=\sum_{n=-1}Tr_{V(n)}\,1 \times q^n,  \hskip2cm  dim V(n)=c(n)  \no
\ea
McKay-Thompson series is given by
\ba
J_g(\tau)=\sum_{n=-1}Tr_{V(n)}\,g \times q^n, \hskip5mm g\in M \no
\ea
where $Tr_{V(n)} \,g$ denotes the character of a group element $g$ in the representation $V(n)$. 
This depends on the conjugacy class $g$ of $M$.  If McKay-Thompson series is known for all conjugacy classes, decomposition of $V(n)$ into irreducible representations become uniquely determined. Series $J_g$ are modular forms with respect to subgroups of $SL(2, Z)$ and 
possess similar properties like the modular J-function such as the genus=0 (Hauptmodul) property. 
 
Phenomenon of monstrous moonshine  has been understood mathematically in early 1990's using the technology of vertex operator algebra. However, we still do not have a 'simple'  physical explanation of this phenomenon.
(A possible connection to heterotic string theory is discussed in a recent article \cite{PPV}).
\section{Mathieu moonshine}

$K_3$ surface :\\

We consider string theory compactified on $K_3$ surface. $K_3$ surface is a complex 2-dimensional 
hyperK\"ahler manifold and ubiquitous in string theory. It possesses 
$SU(2)$ holonomy and  a holomorphic 2-form. 
 Thus the string theory on $K_3$ has an  ${\cal N}$=4 superconformal symmetry with the central charge $c=6$ which contains $SU(2)_{k=1}$ affine symmetry.
 
 Now instead of modular J-function we consider the elliptic genus of $K_3$ surface. Elliptic genus describes the topological invariants of the target manifold and counts the number of BPS states in the theory. Using world-sheet variables it is written as 
 \ba
Z_{elliptic}(z;\tau)=Tr_{{\cal H}_L\times {\cal H}_R}(-1)^{F_L+F_R}e^{4\pi izJ^3_{L,0}}q^{L_0-{c\over 24}}\bar{q}^{\bar{L}_0-{c\over 24}}\no
\ea
 Here $L_0$ denotes the zero mode of the Virasoro operators and $F_L$ and $F_R$ are left and right moving fermion numbers. $J_0^3$ denotes the Cartan generator of affine $SU(2)_1$. In elliptic genus the right moving sector is frozen to the supersymmetric ground states (BPS states) while in the left moving sector all the states in the left-moving Hilbert space ${\cal H}_L$ contribute.
 
 In general it is difficult to compute elliptic genera, however, we were able to evaluate it by 
 making use of Gepner models \cite{EOTY}. Elliptic genus is given by
 \ba
 Z_{K3}(\tau,z)=8\left[\left({\theta_2(\tau,z)\over \theta_2(\tau,0)}\right)^2+
 \left({\theta_3(\tau,z)\over \theta_3(\tau,0)}\right)^2+ \left({\theta_4(\tau,z)\over \theta_4(\tau,0)}\right)^2\right]
\no \ea
 Here $\theta_i(\tau,z)$ are Jacobi theta functions.
 
 We want to see how the Hilbert space ${\cal H}_L$ in elliptic genus decompose into irreducible representations of ${\cal N}$=4 superconformal algebra (SCA).  
 
 Highest weight states of ${\cal N}$=4 SCA are parametrized by the eigenvalues of $L_0$ and $J_0^3$.
 \ba
L_0|h,\ell \rangle=h|h,\ell \rangle,\hskip2cm J_0^3|h,\ell \rangle=\ell|h,\ell \rangle\no
\ea 
 
  There are two different types of representations in $c=6$ SCA.\\
  In the $Ramond$ sector
 \ba
&& \mbox{BPS (massless)  rep.} \hskip3.5cm  h={1\over 4}; \hskip1cm  \ell=0,\,\,{1\over 2}\no\\      
&& \mbox{non-BPS (massive)  rep.} \hskip2.7cm h>{1\over 4}; \hskip1.5cm \ell={1\over 2}\no
\ea
 Character of a representation is defined as
\ba
  Tr_{{\cal R}}(-1)^F q^{L_0}e^{4\pi iz J_0^3} \no
  \ea
 where ${\cal R}$ denotes the representation space.\\
 Index is given by the value of the character at $z=0$,   
 \ba
\mbox{Index}({\cal R})= Tr_{{\cal R}}(-1)^F q^{L_0}\no
 \ea
BPS representations have a non-vanishing index
\ba
&&\mbox{index (BPS, $\ell=0$)}=1\no\\
&&\mbox{index (BPS, $\ell={1\over 2}$)}=-2
\no
\ea
while non-BPS reps. have vanishing indices
\ba
\mbox{index (non-BPS, $\ell={1\over 2}$)}=0. \no
\ea
Characters are given explicitly as \cite{ET}
\ba
ch^{{\small BPS}}_{h={1\over 4},\ell=0}(\tau,z)={\theta_1(z;\tau)^2 \over \eta(\tau)^3}\mu(z;\tau)\no
\ea
where 
\ba
\mu(z;\tau)={-ie^{\pi iz}\over \theta_1(z;\tau)}\sum_n(-1)^n{q^{{1\over 2}n(n+1)}e^{2\pi inz}\over 1-q^ne^{2\pi iz}}\no
\ea  
while non-BPS characters are given by
\ba
ch^{\small non-BPS}_{h,\ell={1\over 2}}=q^{h-{3\over  8}}\hskip2mm{\theta_1(z;\tau)^2\over \eta(\tau)^3}, \hskip1cm h>{1\over 4}\no
\ea
\bigskip
Function $\mu(\tau,z)$ is a typical example of Mock theta function (Lerch sum or Appel function). 
Mock theta functions look like theta functions but they have anomalous modular transformation laws and are difficult to handle. Recently there were developments in understanding the nature of Mock theta functions due to Zwegers\cite{Zwegers}. 

He has introduced a method of regularization which is similar to the ones used in physics. It improves the modular property of Mock theta  functions so that they transform as analytic Jacobi forms.

Now let us make a decomposition of elliptic genus into a sum of characters of ${\cal N}$=4 representaions
 \ba
Z_{K3}(\tau,z)=24ch^{{\small BPS}}_{h={1\over 4},\ell=0}(\tau,z)+2\sum_{n\ge 0}A(n)ch^{{\small non-BPS}}_{h={1\over 4}+n,\ell={1\over 2}}(\tau,z)\no
\ea
At smaller values of $n$, expansion coefficients $A(n)$ may be found by direct series expansion of $Z_{K3}$. 
We find, $A(0)=-1$ and
\bigskip
\begin{equation}
  \label{l1_coefficient_massive}
  \begin{array}{c|rrrrrrrrrr}
     n & 1 & 2 & 3 & 4 & 5 & 6 & 7 & 8  &\cdots\\
    \hline
     A(n)& 45 & 231 & 770 & 2277 & 5796 & 13915 &
     30843 & 65550 & \dots
\nonumber  \end{array}
\end{equation}


Dimensions of some irreducible reps. of Mathieu group $M_{24}$ appear in (1)
\begin{equation}
\begin{array}{crrrrrrrrr}
\mbox{dimensions}:\{ & 45 & 231 & 770 &990&1771& 2024& 2277 & \null \nonumber\\
 &  \null &3312&3520& 5313&  5544& 5796  & 10395 &\cdots\} \nonumber 
\end{array}
\end{equation}
\ba
&& A(6)=13915=3520+10395,\no\\
&& A(7)=30843=10395+5796+5544+5313+2024+1771\no
\ea
\cite{EOT}

$M_{24}$ is a subgroup of $S_{24}$ (permutation group of 24 objects) and  
its order is given by $\approx 10^9$.\\
$M_{24}$ is known for its many interesting arithmetic properties and in particular intimately tied to  the Golay code of efficient error corrections.

\hskip2cm $\mbox{Monster} \supset \mbox{Conway} \supset \mbox{Mathieu} \supset \cdots$

\begin{center}
\section{Mathieu moonshine conjecture}
\end{center}

{
Expansion coefficients of $K_3$ elliptic genus into ${\cal N}$=4 characters are given by the sum of dimensions of representations of Mathieu group $M_{24}$}

We were able to derive analogues of McKay-Thompson series \cite{Gaberdiel, Eguchi-Hikami}.
And then the 
multiplicities $C_R(n)$ of the decomposition of $A(n)$ into representations $R$ 
\ba
A(n)=\sum_R \, C_R(n)\,dim R\no
\ea
were unambiguously determined. 
It tuned out that $C_R(n)$ are all positive integers up to $n\approx 1000$ and this gives a very strong evidence of Mathieu moonshine conjecture.

{\tiny

}

The conjecture is now proved mathematically using the method of mathematical induction.
\cite{Gannon}

Unfortunately the proof so far did not provide much insight into the nature of  Mathieu moonshine. 
The situation looks a bit like the case of monstrous moonshine. 24 of $M_{24}$ will certainly be the Euler number of $K_3$ and $M_{24}$ permutes homology classes. 
There are, however, various indications that string theory on $K_3$ can not have such a high symmetry as $M_{24}$.  Instead of the total Hilbert space the BRS subsector of the theory may possibly possess an enhanced symmetry. It will be interesting to look into the algebraic structures of BPS states to explain Mathieu moonshine.

\vskip1cm

\section{More Moonshine Phenomena}

Mathieu moonshine exists at the intersection of string theory, $K_3$ surface (geometry), (Mock) modular forms and sporadic  discrete symmetry and appears to possess interesting mixture of 
physics and mathematics. Recently there have been intense interests in exploring new types of moonshine phenomena 
other  than Mathieu moonshine. Already several types of new moonshine phenomena have been discovered.  \\

Umbral moonshine  \hskip1cm \cite{Cheng-Duncan-Harvey}\\

 fermions on 24 dim. lattice \hskip4mm \cite{Cheng-Dong-Duncan-Harrison-Kachru-Wrase}\\

 spin 7 manifold \hskip4mm \cite{Benjamin-Harrison-Kchru-Paquette-Whalen}\\


Due to time limitation we discuss only about Umbral moonshine.
Umbral moonshine has a mysterious relationship to the Niemeier lattice. It is known there are 23 (24, if we include  Leech lattice) types  of self-dual lattices in 24 dimensions. It is given by the combination of root lattices of A-D-E type 
together with appropriate weight vectors so that the lattice becomes self-dual.  The simplest examples are
\ba
&& (A_1)^{\,24}  \hskip2cm (k=1)\no \\
 &&(A_2)^{\,12}    \hskip2cm (k=2)    \no \\
 &&(A_3)^{\,8}    \hskip2.2cm (k=3)      \no \\
 &&\,\cdots \hskip3cm \cdots \no
 \ea
 etc.
 If one divides the automorphism groups of Niemeier lattice by the automorphism  group of A-D-E lattice,  
 one obtain isolated discrete groups  
 \ba
 G_k={[\mbox{automorphism group of lattice}]_k\over[\mbox{Weyl group of root lattice}]_k }.\no
\ea 
It turns out that $G_k'$s become the symmetry groups of the Umbral moonshine.
In fact the first one agrees with the Mathieu group $G_1=M_{24}$ and reproduces the Mathieu moonshine.
 The second one $G_2$ agrees with the Mathieu group $M_{12}$ and 
 is assumed to be related to 4-dimensional hyperK\"ahler manifold with $c=12 \hskip1mm (k=2)$.  \\
Analogue of $K_3$ elliptic genus is given by 
\ba
Z(k=2)=4\left[\left({\theta_2(z)\theta_3(z)\over \theta_2(0)\theta_3(0)}\right)^2+\left({\theta_2(z)\theta_4(z)\over \theta_2(0)\theta_4(0)}\right)^2+
\left({\theta_3(z)\theta_4(z)\over \theta_3(0)\theta_4(0)}\right)^2\right]
\no
\ea
By expanding $Z(k=2)$ in terms of characters of representations of $c=12$, ${\cal N}=4$ algebra 
one finds the expansion coefficients decompose into the symmetry group $M_{12}$. 

Here, however, there is something awkward: $Z(k=2)$ does not contain the contribution of vacuum operator ($h=0$ in $NS$ sector) thus the theory appears to describe the geometry of a (singular) non-compact 
four-fold.  The rest of Umbral moonshine series has the same property (absence of identity operator) and their geometrical interpretation is somewhat obscure.

Recently, we have used ${\cal N}=4$ Liouville theory \cite{Eguchi-Sugawara}
which is known to possess some special duality property \cite{SW}. It is possible to embed Umbral series into ${\cal N}=4$ Liouville theory and by using duality we can map Umbral theory at $c=6k$ to its dual theory at $c=6$. 
Thus a Umbral moonshine at $c=6k$ can be mapped to a dual moonshine at $c=6$. We hope this is going to help  geometrical interpretation of Umbral moonshine.

Moonshine symmetries recently discovered in string theory are still very mysterious and we may encounter many more surprises in the near future.\\

{\bf Acknowledgements}
Research of T.E. is supported in part by Japan Ministry of Education, Culture, Sports, Science and Technology under grant Nos. 25400273, 1H5738-1.

\end{document}